\documentclass[a4paper, 10 pt, conference]{ieeeconf}  % Comment this line out
\usepackage{lmodern}  % for bold teletype font
\usepackage{amsmath}  % for \hookrightarrow
\usepackage{amssymb}
\usepackage{listings}
\usepackage[table]{xcolor}
\usepackage{proof}
\usepackage{hhline}
\usepackage{mathabx}

\usepackage{hyperref}
\hypersetup{
    colorlinks=true,
    linkcolor=blue,
    filecolor=magenta,      
    urlcolor=cyan,
}

\IEEEoverridecommandlockouts                              % This command is only
\title{\LARGE \bf
MotePy: A domain specific language for low-overhead machine learning and data processing.

}

\author{Jayaraj Poroor\\
JIFFY.ai\\
{\tt\small jayaraj.poroor@gmail.com}
\thanks{* This work was supported by a grant from Microsoft Research, India.}% <-this % stops a space
}

\numberwithin{equation}{subsection}

\begin{document}

\maketitle

\makeatletter
\def\ps@headings{% default to standard twoside headers, no footers
% will change later if the mode requires otherwise
\let\@oddhead\@empty
\let\@evenhead\@empty
\def\@oddfoot{\footnotesize \hspace{10em} MotePy - A domain specific language for low-overhead machine learning and data processing.)\hfil\thepage}%
\def\@evenfoot{\@IEEEheaderstyle\thepage\hfil\hbox{}}% not used, oneside
}
\def\ps@IEEEtitlepagestyle{% default title page headers, no footers
\let\@oddhead\@empty
\let\@evenhead\@empty
\def\@oddfoot{}%
\let\@evenfoot\@empty
}
\makeatother
%%%%%%%%%%%%%%%%%%%%%%%%%%%%%%%%%%%%%%%%%%%%%%%FooterRight

\pagestyle{headings}
\setcounter{page}{1}
\thispagestyle{IEEEtitlepagestyle}

%%%%%%%%%%%%%%%%%%%%%%%%%%%%%%%%%%%%%%%%%%%%%%%%%%%%%%%%%%%%%%%%%%%%%%%%%%%%%%%%
\begin{abstract}

A domain specific language (DSL), named {\it MotePy} is presented. The DSL offers a high level syntax with low overheads for ML/data processing in time constrained or memory constrained systems. The DSL-to-C compiler has a novel static memory allocator that tracks object lifetimes and reuses the static memory, which we call the {\it compiler-managed heap}.

\end{abstract}

%%%%%%%%%%%%%%%%%%%%%%%%%%%%%%%%%%%%%%%%%%%%%%%%%%%%%%%%%%%%%%%%%%%%%%%%%%%%%%%%
\section{INTRODUCTION}

When writing code for machine learning or data processing applications we want the convenience of a high level syntax of a language like Python \cite{rossum1995python}. Typically this convenience comes with hidden overheads like interpreted execution (e.g., in CPython \cite{roghult2016benchmarking}) or dynamic memory allocation and garbage collection. In many situations these overheads do not pose a concern for the application developer. 

However, in certain situations when we have tight time constraints \cite{liu2000real} or memory constraints \cite{lee2006cyber}, we cannot compromise on the need to have very low overheads. Examples include machine learning pipelines in an OS kernel \cite{negi2005applying} or in small embedded devices \cite{dennis2020edgeml}.

The proposed DSL, {\it MotePy} offers a high level Python-like syntax for specifying modular data processing pipelines and performing vector/matrix operations.

\section{SYNTAX}

Machine learning and data processing applications typically have a pipeline-like execution structure. The main module of a MotePy application is a pipeline specification of the form given in the example below (Figure \ref{fig:samplepipeline}). The pipeline specification is a Python-like list of operations. In the example, the first operation is {\it acquire} function defined in the module {\it datasource} and the second operation is {\it predict} function defined in {\it model} module. MotePy's runtime executes the pipeline in a loop, starting from the first operation, in sequence till the last operation. Once the last operation is complete  the pipeline is re-executed from the beginning, in a loop fashion.

\begin{figure}[h!]
\textbf{pipeline.py}
\begin{lstlisting}
[
datasource.acquire,
model.predict
]
\end{lstlisting}
\caption{A sample pipeline specification}
\label{fig:samplepipeline}
\end{figure}

Figure \ref{fig:datasourcemod} shows the datasource module and Figure \ref{fig:modelmod} shows the model module used in the sample pipeline in Figure \ref{fig:samplepipeline}.
\begin{figure}[h!]
\textbf{datasource.py}
\begin{lstlisting}
data: float[60]

def init() -> void:
    pass

@flow
def acquire() -> void:
    for i in range(0, 60):
        data[i] = float(rand())/RAND_MAX
    printf("Next\n")
    next(data)
\end{lstlisting}
\caption{The datasource module used in the pipeline specification in Figure \ref{fig:samplepipeline}}
\label{fig:datasourcemod}
\end{figure}

\begin{figure}[h!]
\textbf{model.py}
\begin{lstlisting}
N : const int32_t = 60
M : const int32_t = 30
layer0 : float[N][M]
layer1 : float[M]

def random_weights() -> void:
    for j in range(0, M):
        for i in range(0, N):
            layer0[i][j] = 
                float(rand())/RAND_MAX
        layer1[j] = float(rand())/RAND_MAX

def init() -> void:
    random_weights()

@flow
def predict(features: float[N]) -> void:
    res: float
    tmp : float[30]
    printf("Prediction\n")
    tmp = features * layer0
    res = tmp * layer1
    #res = exp(res)
    printf("Result %f\n", res)
\end{lstlisting}
\caption{The model module used in the pipeline specification in Figure \ref{fig:samplepipeline}}
\label{fig:modelmod}
\end{figure}

Though the syntax is close to Python, the language differs from Python in its implementation in significant ways:
\begin{itemize}
    \item MotePy is statically typed. Its type annotation syntax is based on Python's optional type annotation syntax, but uses MotePy's own notation similar to that of Java for array typing.The reason for own notation for array typing is that type annotations are not common in Python and Java's array notation is familiar to most programmers.
    \item MotePy supports only static memory allocation. Only primitive types are allocated on the stack. Rest are allocated in the static data area. However, unlike other languages such as C or C++ the MotePy compiler tracks lifetimes of array objects and reuses memory. We call this the {\it compiler-managed heap}.
\end{itemize}

Every MotePy module has an init function and a special {\it flow} function. The flow function is the operation invoked as part of the pipeline execution and is decorated with {\it @flow} syntax. Apart from these two functions a module may define any number of functions to perform its function.

In Figure \ref{fig:datasourcemod} the data array is randomly initialized and then passed onto the next stage in the pipeline by invoking a special {\it next} function. The next function enables us to specify at what point in the execution the next stage in pipeline must be executed. If a flow function returns without invoking the next function the pipeline is re-executed from the beginning. For instance, if the data acquisition is not complete, then the acquisition stage may return without calling next. The MotePy runtime will automatically re-execute the pipeline.

Figure \ref{fig:modelmod} shows a neural-network-like processing stage. The network has two layers of weights that are randomly initialized in the example. In an actual system the weights will be loaded from a file. The code shows high level syntax for vector/matrix operations, which are translated by the MotePy compiler into low-level loop operations. 

The code also shows invoking the C function {\it printf}. MotePy allows us to invoke C/C++ functions from MotePy code without any need for a special foreign function calling interface. The MotePy array/matrix layout is same as that of C/C++ enabling us to pass arrays/matrices back and forth between C/C++ code without any transformation overheads.

The MotePy compiler translates the high level DSL code to low level C/C++ code, which is then compiled using a platform C/C++ compiler such as GCC to get the executable.

\section{COMPILER-MANAGED HEAP}

Figure \ref{fig:samplepipeline2} shows a two-stage MotePy pipeline. Stage 1 initializes vec1, does some computations on it and then invokes next for Stage 2 processing. Stage 2 initializes vec2, does some computations on it, then initializes vec3 and does some computations on it.

Compiling and running the program gives the following output:

{\bf \tt \ \\
Address of vec1: 94473766309984\\
Address of vec2: 94473766309984\\
Address of vec3: 94473766309984\\}

The output shows that the addresses of vec1, vec2, and vec3 are same\footnote{The generated executable has an optional parameter for specifying the number of pipeline iterations. In this case a value of `1' was passed. If none is passed, the pipeline is executed in an infinite loop.}. The reason is that the compiler allocates static memory for the vectors and reuses the memory based on the variable lifetimes. It may be noted that owing to Address Space Layout Randomization\cite{jang2016aslr} the printed addresses will be different during each program execution.

In the case of this example, the life time of vec1 is in the program region from line numbers 8-11 in stage1.py. Life time of vec2 is from line numbers 8 - 11 in stage2.py and that of vec3 is from line numbers 12 - 15 in stage2.py. Since the lifetimes are non-overlapping the compiler intelligently reuses the static memory area for the three vectors.

\begin{figure}[h!]
\textbf{pipeline.py}
\begin{lstlisting}
[
stage1.process,
stage2.process
]
\end{lstlisting}

\textbf{stage1.py}
\begin{lstlisting}
1. vec1: float[60]
2. 
3. def init() -> void:
4.    #vec1 = 0
5.    pass
6.
7. def process() -> void:
8.    vec1 = 0 #initialization
9.    vec1 = vec1 + vec1 #some processing
10.   printf("Address of vec1: %ld\n", 
11.           uint64_t(vec1))
12.   next()
\end{lstlisting}

\textbf{stage2.py}
\begin{lstlisting}
1. vec2: float[60]
2. vec3: float[60]
3. 
4. def init() -> void:
5.   pass
6.
7. def process() -> void:
8.    vec2 = 0 #initialization
9.    vec2 = vec2 + vec2
10.   printf("Address of vec2: %ld\n", 
11.         uint64_t(vec2))
12.   vec3 = 0
13.   vec3 = vec3 + vec3
14.   printf("Address of vec3: %ld\n", 
15.         uint64_t(vec3))
\end{lstlisting}

\caption{A two-stage MotePy pipeline}
\label{fig:samplepipeline2}
\end{figure}

Figure \ref{fig:stage1redef} shows a modified version of the Stage 1 of the pipeline. Here the vec1 is initialized in the init function. The process function computes a new value of vec1 based on the current value. Compiling and running the pipeline with this modified Stage 1 gives the following result:

{\bf \tt \ \\
Address of vec1: 94035385520224\\
Address of vec2: 94035385520464\\
Address of vec3: 94035385520464\\
}

The output shows that the compiler allocates a separate non-overlapping memory area for vec1 and uses the same memory area for vec2 and vec3. The compiler's allocator now knows that vec1's lifetime overlaps with that of vec2 and vec3 and hence allocates separate memory for vec1.

\begin{figure}[h!]
\textbf{stage1.py}
\begin{lstlisting}
1. vec1: float[60]
2. 
3. def init() -> void:
4.    vec1 = 0
5.    #pass
6.
7. def process() -> void:
8.    #vec1 = 0 #initialization
9.    vec1 = vec1 + vec1 #some processing
10.   printf("Address of vec1: %ld\n", 
11.           uint64_t(vec1))
12.   next()
\end{lstlisting}

\caption{A modified version of Stage 1}
\label{fig:stage1redef}
\end{figure}

MotePy's static allocator allocates along the 2-dimensional space with address as one axis and variable lifetime as the other axis. In comparison, the static allocators of languages like C/C++ simply allocates along the address line.

The objective of the MotePy allocator is to minimize the allocated space subject to the constraint that no two blocks overlap in this 2-dimensional space. The current implementation uses a greedy approach, which may yield sub-optimal results. More work on the allocator is planned in the future.

MotePy compiler code, along with sample applications is released in open source and is available at \url{https://github.com/jayarajporoor/motepy}.

\section{ACKNOWLEDGEMENTS}

I would like to thank Microsoft Research, India for the grant support. I am especially thankful to Dr. Sriram Rajamani, Satish Sangameswaran, and Dr. Harsha Vardhan Simhadri for the support during the course of this work. I thank Nandagopal, Sandesh Ghanta, and Surya Chaitanya who worked with me as student interns during the initial development of this work.
\bibliographystyle{unsrt}

\end{document}